\documentclass[%
preprint,
 amsmath,amssymb,
 aps,
]{revtex4-1}

\usepackage{color}
\usepackage{graphicx}
\usepackage{dcolumn}
\usepackage{bm}
\usepackage{hyperref}

\def\sub#1{_{\mathrm{#1}}}
\def\up#1{^{\mathrm{#1}}}
\def\Vec#1{\boldsymbol #1}

\newcommand {\beq}{\begin{eqnarray}}
\newcommand {\eeq}{\end{eqnarray}}

\begin{document}


\title{Interpolating relativistic and nonrelativistic Nambu--Goldstone and Higgs modes 
}

\author{Michikazu Kobayashi$^1$, Muneto Nitta$^{2}$}
\affiliation{%
$^1$Department of Physics, Kyoto University, Oiwake-cho, Kitashirakawa, Sakyo-ku, Kyoto 606-8502, Japan \\
$^2$Department of Physics, and 
Research and Education Center for Natural Sciences, Keio University, Hiyoshi 4-1-1, Yokohama, Kanagawa 223-8521, Japan
}%

\date{\today}

\begin{abstract}
When a continuous symmetry is spontaneously broken 
in nonrelativistic theories, 
there appear  Nambu--Goldstone (NG) modes,
the dispersion relations of which are either 
linear (type I) or 
quadratic (type II).  
We give a general framework to
interpolate between
relativistic and nonrelativistic NG modes, 
revealing a nature of type-I and -II NG modes 
in nonrelativistic theories.
The interpolating Lagrangians have the 
nonlinear Lorentz invariance which 
reduces to the Galilei or Schr\"odinger 
invariance in the nonrelativistic limit.
We find that type-I and type-II NG modes 
in the interpolating region are accompanied with 
a Higgs mode and a chiral NG partner, respectively,  
both of which are gapful.
In the ultrarelativistic limit, 
a set of a type-I NG mode and its Higgs partner remains,
while a set of a type-II NG mode and 
its gapful NG partner turns to  a set of two type-I NG modes.
In the nonrelativistic limit, 
the both types of accompanied gapful modes 
become infinitely massive, disappearing from the spectrum.
The examples contain a phonon in Bose--Einstein condensates 
or helium superfluids, 
a phonon and magnon in spinor Bose--Einstein condensates, 
a magnon in ferromagnets,  
and a kelvon and dilaton-magnon localized around 
a Skyrmion line in ferromagnets.

\end{abstract}


\maketitle

\section{Introduction}

When a continuous symmetry is spontaneously broken 
in nonrelativistic theories, 
there appear  Nambu--Goldstone (NG) modes,
the dispersion relations of which are either 
linear (type I) or 
quadratic (type II).  
The numbers of type-I and -II NG modes satisfy
the Nielsen--Chadha inequality \cite{Nielsen:1975hm}.
After the crucial observation 
\cite{Nambu:2004yia}, 
the numbers of type-I and -II NG modes were summarized 
as the Watanabe--Brauner relation 
between those numbers and the rank of a matrix 
consisting of 
the commutation relations of broken generators 
evaluated in the ground state 
\cite{Watanabe:2011ec}. 
The relation classifies types A and B instead of types I and II.
This  relation was proved recently
in the effective theory approach \cite{Watanabe:2012hr,Watanabe:2014fva},   
the Mori projection method \cite{Hidaka:2012ym}
and later by the Bogoliubov theory \cite{Takahashi:2014vua}.

In the presence of topological solitons or defects, 
there appear NG modes localized around them 
such as translational zero modes. 
When NG modes are normalizable such as 
a domain wall \cite{Kobayashi:2014xua} 
and Skyrmion line \cite{Watanabe:2014pea,Kobayashi:2014eqa} 
in ferromagnets, localized type-B NG modes have 
usual quadratic dispersion relations and are of type II. 
On the other hand, when 
NG modes are non-normalizable such as 
a domain wall 
in two-component Bose--Einstein condensates (BECs) 
and a vortex in scalar BECs or $^4$He superfluid, 
type-B NG modes have 
usual quadratic dispersion relations and are of type II 
when the transverse system size is small compared with 
the wavelength of NG modes 
(see, e.g.~Refs.~\cite{Kobayashi:2013gba,Takahashi:2014vua}), 
but they have noninteger dispersion relations 
in infinite system sizes 
\cite{Takeuchi:2013mwa,Watanabe:2014zza,Takahashi:2014vua}.
The formulas of dispersion relations 
interpolating small and large system sizes 
were recently obtained in Ref.~\cite{Takahashi:2015caa}.
A relationship between the number of NG modes and the homotopy group for topological solitons was also studied \cite{Higashikawa:2015}.

Other developments include, for instance,  
space-time symmetry breaking \cite{Hidaka:2014fra}, 
gauge symmetry breaking accompanied 
with the Higgs mechanism 
\cite{Hama:2011rt,Gongyo:2014sra,Watanabe:2014qla}, 
finite temperature and density \cite{Hayata:2014yga}, 
topological interaction \cite{Brauner:2014ata}, 
and quasi-NG modes \cite{Nitta:2014jta}.

\begin{table}[h]
\begin{tabular}{|c|c|c|c|c|}
\hline
& Parameters & Symmetry & Type-I NG mode 
& Type-II NG mode \\
\hline
Ultrarelativistic  & $\mu \to 0$       & Lorentz &
   1 type-I + 1 Higgs & 2 type-I  \\ 
Relativistic & $0<c,\mu < \infty$& Lorentz & 
   1 type-I + 1 Higgs & 1 type-II + 1 gapped \\ 
Nonrelativistic   & $c \to \infty$  & Galilei (Shr\"odinger) & 
   1 type-I & 1 type-II \\ \hline
\end{tabular}
\caption{Interpolation of 
type-I and type-II NG modes between 
nonrelativistic and ultrarelativistic theories. 
}
\label{table}
\end{table}

In general,
it is usually said that 
only type-I NG modes are possible in 
Lorentz invariant theories.
It is, however, not yet clear how NG modes are 
interpolated between relativistic and nonrelativistic 
theories, in particular, how type-II NG modes 
in nonrelativistic theories reduce to 
type-I NG modes in relativistic theories 
when both the theories are interpolated.
In this paper, we clarify 
how NG modes are 
interpolated between relativistic and nonrelativistic 
theories, as summarized in Table \ref{table}. 
We first consider relativistic Lagrangians and introduce 
a chemical potential for particles. 
The resulting Lagrangians, 
containing both first- and second-order time derivative terms, 
interpolate relativistic and nonrelativistic 
Lagrangians in the two limits: 
the second time derivative vanishes 
in the nonrelativistic limit $c\to \infty$
with the speed $c$ of the light, 
while  the first time derivative vanishes 
in the relativistic limit $\mu \to 0$,  
in which the chemical potential is sent to zero.
The latter case is often called ultrarelativistic, 
so we use this terminology because 
the Lorentz invariance exists in the intermediate region.  
We first point out that interpolating Lagrangians 
have the nonlinear Lorentz invariance, 
which reduces to the Galilei or Schr\"odinger 
invariance in the nonrelativistic limit $c\to \infty$. 
The Lorentz invariance becomes manifest in 
the ultrarelativistic limit $\mu \to 0$.
We find that 
there can exist either 
a type-I or type-II NG mode in the intermediate Lagrangian,
even in the presence of the Lorentz invariance. 
Correspondingly, the Watanabe--Brauner relation holds 
in the intermediate region even in the presence of the 
Lorentz invariance; 
a commutator of two generators does not vanish 
for a type-II NG mode. 
We also find that 
each of either the type-I or type-II mode is accompanied with
a gapful mode.
The gapful mode  accompanied with 
 a type-I NG mode can be identified with a Higgs mode, 
while that accompanied with 
 a type-II NG mode  can be referred as a ``chiral partner" or a 
``gapful NG partner".
In the ultrarelativistic limit, 
a set of a type-I NG mode and its Higgs partner remains 
as it is,
while a set of a type-II NG mode and 
gapful NG partner becomes  a set of two type-I NG modes.
On the other hand, in the nonrelativistic limit, 
both the Higgs partner of a type-I NG mode 
and gapful NG parter of a type-II NG mode 
become infinitely massive 
and disappear from the spectrum, 
and there remains only the type-I or type-II NG mode.
This mechanism reveals a nature of type-I and II NG modes.
We show these in typical examples of 
both bulk NG modes and NG modes localized around 
a topological soliton.
The bulk examples contain a phonon in scalar BECs and 
a magnon in ferromagnets, while 
soliton examples contain  
a kelvon and dilaton-magnon localized around 
a Skyrmion line in isotropic ferromagnets.
An another example is a ripplon-magnon localized 
around a domain wall in anisotropic ferromagnets 
studied in Ref.~\cite{Kobayashi:2014xua}.
In all examples, we give two approaches: 
the effective theory 
and linear response theory (the Bogoliubov--de Gennes equations).

The interpolating Lagrangians 
containing both first- and second-order time derivatives
that we consider 
in this paper appear in various contexts of 
both theoretical and experimental physics,
and thereby our results yield suggestions of several theoretical and experimental works.
As we denoted above, they describe
relativistic field theories with the finite chemical potential. 
In the quantum mechanical framework, it is well known that these models naturally give the complex probabilities in the path-integral formalism as long as the chemical potential is finite \cite{Parisi:1983,Aarts:2008rr}, and we can expect some qualitative change from zero to finite chemical potentials.
Even in the semiclassical framework for symmetry-broken systems, our results show that there is a drastic change of low-energy modes both in bulk and topological defects: 
the coupling of the type-I NG mode and Higgs mode or the coupling of two type-I NG modes to one type-II NG mode.
The interpolation between the relativistic and nonrelativistic frameworks appears in various ultracold atomic systems.
Recently, the existence of Higgs mode in strongly interacting lattice bosons close to the superfluid-Mott insulating transition point has been theoretically \cite{Altman:2002,Altman:2008,Pollet:2012,Grass:2011,Krutitsky:2011,Nakayama:2015} and experimentally \cite{Bissbort:2011} proposed and confirmed. 
In this system,
 the first-order time derivative term is prohibited, and the second-order time derivative term becomes important at the transition point because of the particle-hole symmetry, by which the Higgs mode can be expected.
In the superfluid phase far from the transition point, it is well known that the first-order time derivative term is dominant and the Higgs mode is absent.
Our results explain how the Higgs and NG modes are changed as the parameter changes to the transition point.
We show two two examples.
One is the charged fermionic systems close to the BEC-BCS crossover point \cite{Liu:2015} 
that has been theoretically predicted to contain the both first- and second-ordered time derivative terms, 
predicting the existence of the Higgs mode.
The other example is the magnetic model \cite{Beekman:2014}.
As already known, two type-I NG modes exist in antiferromagnets, while one type-II NG mode exists in ferromagnets.
In a canted ferromagnet between the ferromagnet and antiferromagnet, it has been reported that there appear one type-II NG mode and one gapful Higgs mode, which is quite similar to what we obtain in the interpolating region between relativistic and nonrelativistic models even though the magnetic model itself is nonrelativistic.

This paper is organized as follows.
In Sec.~\ref{sec:bulk}, we discuss bulk NG modes.
We study phonons in a scalar BEC and 
magnons in ferromagnets 
as examples of type-I and -II NG modes 
in Secs.~\ref{subsec-phonon-BEC} and 
\ref{sec:magnon}, respectively. 
In Sec.~\ref{sec:soliton}, we discuss NG modes localized around 
topological solitons. We study kelvons and dilaton-magnons 
localized around a Skyrmion line in ferromagnets 
as examples of type-II NG modes. 
Section~\ref{sec:summary} is devoted to a summary and discussion.
In Appendix \ref{sec:spinor}, we give a further example 
of a spinor BEC that contains both type-I and -II NG modes 
in the bulk.

\section{NG modes in the bulk}\label{sec:bulk}
\subsection{Interpolating type-I NG mode: 
Phonons in scalar BEC} \label{subsec-phonon-BEC}
Let us consider the Lagrangian density for 
a single complex scalar field $\psi$, 
interpolating relativistic and nonrelativistic theories,
\begin{align}
\begin{split}
& \mathcal{L} = \frac{|\partial_t \psi|^2}{c^2} + i \mu (\psi^\ast \partial_t \psi - \partial_t \psi^\ast \psi) - |\nabla \psi|^2 - \frac{g}{2} (|\psi|^2 - \rho)^2,
\label{eq-scalar-Lagrangian}
\end{split}
\end{align}
where $g$ is the coupling constant and 
$\rho$ is the real positive constant giving 
a vacuum expectation value. 
This Lagrangian density interpolates between two extreme cases;
it reduces to the relativistic Goldstone model
in the ultrarelativistic limit $\mu \to 0$ 
and to the Gross--Pitaevskii model in 
the nonrelativistic limit $c \to \infty$.
In the generic region, the Lagrangian density
is invariant under the Lorentz transformation  
\begin{align}
\begin{split}
& t^\prime = \gamma \bigg(t - \frac{v x}{c^2} \bigg), \quad 
x^\prime = \gamma (x - v t), \quad
\partial_{t^\prime} = \gamma \partial_t + \gamma v \partial_x, \quad 
\partial_{x^\prime} = \frac{\gamma v}{c^2} \partial_t + \gamma \partial_x, \\
& \psi^\prime = e^{i \mathcal{S}} \psi, \quad 
\mathcal{S} = - \mu c^2 \bigg\{ (1 - \gamma) t + \frac{\gamma v x}{c^2} \bigg\}, \quad 
\gamma = \frac{1}{\sqrt{1 - v^2 / c^2}},
\label{eq:Lorentz}
\end{split}
\end{align}
which reduces to the Galilei or Schr\"odinger transformation 
in the nonrelativistic limit $c \to \infty$.
The equation of motion for $\psi^\ast$ reads
\begin{align}
\begin{split}
- \frac{1}{c^2} \partial^2_t \psi + 2 i \mu \partial_t \psi = - \nabla^2 \psi + g (|\psi|^2 - \rho) \psi, \label{eq-GP-equation}
\end{split}
\end{align}
which has the static constant solution $\psi_0 = \sqrt{\rho}$.

\subsubsection{Low-energy effective theory}

The low-energy dynamics around the static solution $\psi_0$ can be discussed by considering the low-energy effective theory.
We introduce fluctuations of amplitude $f(\Vec{x},t)$ and phase $\theta(\Vec{x},t)$ around $\psi_0$:
\begin{align}
\psi = \psi_0 \{ 1 + f(\Vec{x},t) \} e^{i \theta(\Vec{x},t)}.
\label{eq-scalar-effective-theory}
\end{align}
Inserting Eq. \eqref{eq-scalar-effective-theory} into Eq. \eqref{eq-scalar-Lagrangian}, we obtain the effective Lagrangian density
\begin{align}
\frac{\mathcal{L}\sub{eff}}{\rho} = \frac{\dot{f}^2 + \dot{\theta}^2}{c^2} - 2 \mu (1 + 2 f) \dot{\theta} - (\nabla f)^2 - (\nabla \theta)^2 - 2 g \rho f^2 + O((f,\theta)^3).
\end{align}
The low-energy dynamics of $f$ and $\theta$ derived from the Euler--Lagrange equations reads
\begin{align}
\frac{\ddot{f}}{c^2} + 2 \mu \dot{\theta} - \nabla^2 f + 2 g \rho f = 0, \quad
\frac{\ddot{\theta}}{c^2} - 2 \mu \dot{f} - \nabla^2 \theta = 0.
\label{eq-scalar-Euler-Lagrange}
\end{align}
Typical solutions of Eq. \eqref{eq-scalar-Euler-Lagrange} are $f = f_0 \cos(\Vec{k} \cdot \Vec{x} - \omega t + \delta)$, $\theta = \theta_0 \sin (\Vec{k} \cdot \Vec{x} - \omega t + \delta)$ with the dispersions
\begin{align}
\begin{split}
& \omega\up{H}_\pm = \pm c \sqrt{2 \mu^2 c^2 + k^2 + g \rho + \sqrt{4 \mu^4 c^4 + 4 \mu^2 c^2 (k^2 + g \rho) + g^2 \rho^2}}, \\
& \omega\up{NG}_\pm = \pm c \sqrt{2 \mu^2 c^2 + k^2 + g \rho - \sqrt{4 \mu^4 c^4 + 4 \mu^2 c^2 (k^2 + g \rho) + g^2 \rho^2}}.
  \label{eq:higgs-phonon}
\end{split}
\end{align}
In the long-wavelength limit for $k \to 0$, 
these dispersions reduce to
\begin{align}
\begin{split}
\omega\up{H}_\pm = \pm c \sqrt{4 \mu^2 c^2 + 2 g \rho} + O(k^2), \quad
\omega\up{NG}_\pm = \pm c k \sqrt{\frac{g \rho}{2 c^2 \mu^2 + g \rho}} + O(k^2)
\end{split}
\end{align}
giving rise to one gapful ($\omega\sub{H}$) and one gapless ($\omega\sub{NG}$) mode, 
identified as Higgs and NG modes, 
respectively.
The amplitudes $f_0$ for $\omega\up{H}_\pm$ and $\omega\up{NG}_\pm$ are obtained as 
\begin{align}
\begin{split}
& f\up{H}_{0 \pm} = \mp \frac{\theta_0 \sqrt{2 \mu^2 c^2 + g \rho}}{\sqrt{2} \mu c} + O(k^2), \quad
f\up{NG}_{0 \pm} = \pm \frac{\theta_0 \mu c k}{\sqrt{g \rho (2 \mu^2 c^2 + g^2 \rho^2)}} + O(k^2). \label{eq-scalar-amplitude}
\end{split}
\end{align}
This is the relationship between amplitudes of fluctuations for coupled dynamics of $f$ and $\theta$.
In the long-wavelength limit $k \to 0$, we obtain $f\up{NG}_{0\pm} \to 0$, indicating that the oscillation of $f$ vanishes and there remains the oscillation of $\theta$ as the pure phase mode.

In the ultrarelativistic limit $\mu \to 0$, the dynamics of $f$ and $\theta$ in Eq. \eqref{eq-scalar-Euler-Lagrange} are independent of each other, and amplitudes $f_0$ and $\theta_0$ become independent variables with dispersions
\begin{align}
\omega\up{H}_\pm \to \pm c \sqrt{k^2 + 2 g \rho}, \quad
\omega\up{NG}_\pm \to \pm c k.
\end{align}
The gapful dispersion $\omega\up{H}_\pm$ reduces to that for $f$, while the gapless dispersion $\omega\up{NG}_\pm$ reduces that for $\theta$, i.e., the Higgs and NG modes get to pure amplitude and phase oscillations respectively.
In the nonrelativistic limit $c \to \infty$, the Higgs mode disappears with the divergent dispersion $\omega\up{H}_\pm \to \infty$. 
The NG modes remain coupled oscillations of $f$ and $\theta$ with
\begin{align}
\begin{split}
\omega\up{NG}_\pm \to \pm \frac{k \sqrt{k^2 + 2 g \rho}}{2 \mu}, \quad
f\up{NG}_{0 \pm} \to \pm \frac{\theta_0 k}{\sqrt{k^2 + 2 g \rho}}.
\end{split}
\end{align}
In the long-wavelength limit $k \to 0$, the NG mode is always the pure phase mode with arbitrary $\mu$.

\subsubsection{Linear-response theory}

In the linear-response framework, the dynamics of $\psi$ can be written as 
$\psi \to \psi_0 + u + v^\ast$ with fluctuations $u$ and $v^\ast$.
Here, rewriting $u$ and $v$ as $u = u_0 e^{i (\Vec{k} \cdot \Vec{x} - \omega t + \delta)}$, $v = v_0 e^{i (\Vec{k} \cdot \Vec{x} - \omega t + \delta)}$ with $u_0,\ v_0 \in \mathbb{R}$, and inserting $\psi$ into Eq. \eqref{eq-GP-equation}
leads to the Bogoliubov equation 
\begin{align}
\begin{pmatrix}
\omega^2 / c^2 + 2 \mu \omega - k^2 - g \rho & - g \rho \\
- g \rho & \omega^2 / c^2 - 2 \mu \omega - k^2 - g \rho
\end{pmatrix}
\begin{pmatrix} u_0 \\ v_0 \end{pmatrix} + O((u_0,v_0)^2) = 0
\end{align}
with the dispersion relation $\omega = \omega_\pm\up{H,NG}$.
The fluctuation $\delta \psi = u + v^\ast$ becomes
\begin{align}
\delta \psi \propto \cos(\Vec{k} \cdot \Vec{x} - \omega_\pm\up{H} t + \delta) + \frac{2 \mu^2 c^2 \pm \mu c \sqrt{4 \mu^2 c^2 + 2 g \rho}}{g \rho} e^{- i (\Vec{k} \cdot \Vec{x} - \omega_\pm\up{H} t + \delta)} + O(k^2),
\label{eq-scalar-Higgs-Bogoliubov}
\end{align}
for Higgs modes with the dispersion $\omega\up{H}_\pm$, and
\begin{align}
\delta \psi \propto i \sin(\Vec{k} \cdot \Vec{x} - \omega_\pm\up{NG} t + \delta) \pm \frac{\mu c k}{\sqrt{g \rho (2 \mu^2 c^2 + g \rho)}} e^{- i (\Vec{k} \cdot \Vec{x} - \omega_\pm\up{NG} t + \delta)} + O(k^2),
\label{eq-scalar-NG-Bogoliubov}
\end{align}
for NG modes with the dispersion $\omega\up{NG}_\pm$.
Because the ground state $\psi_0 = \sqrt{\rho}$ has only the real part, the first terms of the right-hand sides in Eqs.~\eqref{eq-scalar-Higgs-Bogoliubov} and \eqref{eq-scalar-NG-Bogoliubov} can be regarded as the amplitude and phase oscillations respectively.
Both the second terms of the right-hand sides  in Eqs.~\eqref{eq-scalar-Higgs-Bogoliubov} and \eqref{eq-scalar-NG-Bogoliubov} are coupled oscillations of the amplitude and phase.
They vanish in the ultrarelativistic limit, which reveals that the Higgs and NG modes become purely 
amplitude and phase oscillations, respectively.
In the nonrelativistic limit, on the other hand, the NG mode
\begin{align}
\delta \psi \propto i \sin(\Vec{k} \cdot \Vec{x} - \omega_\pm\up{NG} t + \delta) \pm \frac{k}{\sqrt{2 g \rho}} e^{- i (\Vec{k} \cdot \Vec{x} - \omega_\pm\up{NG} t + \delta)} + O(k^2),
\end{align}
remains to be a coupled amplitude and phase oscillation.
In the long-wavelength limit $k \to 0$, the NG mode is always the pure phase oscillation with arbitrary $\mu$.
Equations.~\eqref{eq-scalar-Higgs-Bogoliubov} and \eqref{eq-scalar-NG-Bogoliubov} are consistent with the expansion of $\psi = \sqrt{\rho} \{ 1 + f_{0\pm}\up{H,NG} \cos(\Vec{k} \cdot \Vec{x} - \omega_\pm\up{H,NG} t + \delta) \} \exp\{ i \theta_0 \sin (\Vec{k} \cdot \Vec{x} - \omega_\pm\up{H,NG} t + \delta) \}$ around $\theta_0 = 0$, which reveals that both the low-energy effective theory and linear-response theory give the same Higgs and NG modes.
Similar behaviors of NG and Higgs modes are theoretically reported for strongly interacting lattice bosons close to superfluid-Mott insulating transition point \cite{Grass:2011,Krutitsky:2011,Nakayama:2015}.

\subsection{Interpolating type-II NG mode: Magnons in ferromagnets}
\label{sec:magnon}
We consider the interpolating Lagrangian density  
for the continuum Heisenberg model or 
the $O(3)$ nonlinear sigma (${\mathbb C}P^1$) model
\begin{align}
\begin{split}
\mathcal{L}
&= \frac{|\dot{u}|^2}{c^2 (1 + |u|^2)^2} + \frac{i \mu (u^\ast \dot{u} - \dot{u}^\ast u)}{1 + |u|^2} - \frac{|\nabla u|^2}{(1 + |u|^2)^2},
\end{split} \label{eq-uniform-CP1}
\end{align}
where $u \in \mathbb{C}$ is the complex projective coordinate of $\mathbb{C}P^1$, defined as $\phi^T = (1, u)^T / \sqrt{1 + |u|^2}$ with normalized two complex scalar fields 
$\phi = (\phi_1, \phi_2)^T$ with $|\phi_1|^2+|\phi_2|^2=1$.
This Lagrangian density is invariant 
under the following Lorentz transformation:
\begin{align}
\begin{split}
& t^\prime = \gamma \bigg(t - \frac{v x}{c^2} \bigg), \quad
x^\prime = \gamma (x - v t), \quad
\partial_{t^\prime} = \gamma \partial_t + \gamma v \partial_x, \quad
\partial_{x^\prime} = \frac{\gamma v}{c^2} \partial_t + \gamma \partial_x, \\
& u^\prime = e^{i \mathcal{S}} u, \quad
\partial_t \mathcal{S} = - (1 - \gamma) \mu c^2 (1 + |u|^2), \quad \partial_x \mathcal{S} = - \gamma \mu v (1 + |u|^2).
\end{split}
\end{align}
This reduces to the Galilei or Schr\"odinger transformation in 
the nonrelativistic limit $c\to \infty$.
The equation of motion for $u$ reads
\begin{align}
\frac{(|u|^2 + 1) \ddot{u} - 2 u^\ast \dot{u}^2}{c^2} - 2 i \mu (|u|^2 + 1) \dot{u} = (|u|^2 + 1) \nabla^2 u - 2 u^\ast (\nabla u)^2, \label{eq-magnetic-GP}
\end{align}
which has the uniform and static solution $u_0 = \mathrm{const}$.

Under the Hopf map for a 3-vector of real scalar fields $\Vec{n} = \phi^\dagger \Vec{\sigma} \phi$ with the Pauli matrices $\Vec{\sigma}$, the Lagrangian density  \eqref{eq-uniform-CP1} describes the isotropic Heisenberg ferromagnets
\begin{align}
\begin{split}
\mathcal{L}
&= \frac{1}{c^2} \bigg\{ \frac{|\dot{\Vec{n}}|^2}{4} + \frac{\mu c^2 (\dot{n}_1 n_2 - n_1 \dot{n}_2)}{1 + n_3} \bigg\} - \frac{|\nabla \Vec{n}|^2}{4}.
\end{split}
\end{align}

\subsubsection{Low-energy effective theory}

Here, we consider the low-energy effective theory for the low-energy excitation,
with fixing a uniform and static solution $u_0 = 0$ and its fluctuation $\delta u = \alpha + i \beta$ with $\alpha,\ \beta \in \mathbb{R}$.
In terms of $\Vec{n}$, $u_0 = 0$ is equivalent to  $n_3 = 1$, and $\alpha$ and $\beta$ are the fluctuations of $n_1$ and $n_2$, respectively.
The effective Lagrangian density becomes
\begin{align}
\begin{split}
\mathcal{L}\sub{eff}
&= \frac{\dot{\alpha}^2 + \dot{\beta}^2}{c^2}  + 2 \mu (\dot{\alpha} \beta - \alpha \dot{\beta}) - (\nabla \alpha^2 + \nabla \beta^2) + O((\alpha, \beta)^3).
\label{eq-magnon-effective-Lagrangian}
\end{split}
\end{align}
The low-energy dynamics of $\alpha$ and $\beta$ becomes
\begin{align}
\begin{split}
\frac{\ddot{\alpha}}{c^2} + 2 \mu \dot{\beta} - \nabla^2 \alpha = 0, \quad
\frac{\ddot{\beta}}{c^2} - 2 \mu \dot{\alpha} - \nabla^2 \beta. \label{eq-magnon-dynamics}
\end{split}
\end{align}
As in the previous case for phonons in a scalar BEC, the dynamics of $\alpha$ and $\beta$ are independent of each other only in the ultrarelativistic limit $\mu \to 0$.
Typical solutions are $\alpha = \alpha_{0} \cos(\Vec{k} \cdot \Vec{x} - \omega t + \delta)$, $\beta = \beta_{0} \sin(\Vec{k} \cdot \Vec{x} - \omega t + \delta)$, with
\begin{align}
\begin{split}
& \omega\up{H}_\pm = \pm c \Big( \sqrt{k^2 + \mu^2 c^2} + \mu c \Big) = \pm \bigg( 2 \mu c^2 + \frac{k^2}{2 \mu} \bigg) + O(k^4), \quad \alpha\up{H}_{0\pm} = \mp \beta\up{H}_{0\pm}, \\
& \omega\up{NG}_\pm = \pm c \Big( \sqrt{k^2 + \mu^2 c^2} - \mu c \Big) = \pm \frac{k^2}{2 \mu} + O(k^4), \quad \alpha\up{NG}_{0\pm} = \pm \beta\up{NG}_{0\pm}.
 \label{eq:dispersion-magnon}
\end{split}
\end{align}
The second solution is a type-II NG mode which is a magnon, 
and the first one is its chiral massive partner which we may call a ``massive magnon''. 
In the ultrarelativistic limit $\mu \to 0$, 
Eq. (\ref{eq:dispersion-magnon})  
reduces to $\omega\up{H,NG}_\pm = \pm c k$ and $\alpha\up{H,NG}_{0 \pm}$ and $\beta\up{H,NG}_{0\pm}$ are independent of each other,
which implies that the type-II NG and Higgs modes reduce to two type-I NG modes. 
In the nonrelativistic limit $c \to \infty$, on the other hand, 
it reduces to 
\begin{align}
\begin{split}
& \omega\up{H}_\pm \to \infty, \quad \omega\up{NG}_\pm \to \pm \frac{k^2}{2 \mu}.
\end{split}
\end{align}
While the type-II NG mode remains gapless, 
the chiral massive partner becomes infinitely massive 
and disappears from the spectrum.

\subsubsection{Linear-response theory}

We consider the dynamics of magnons in 
the linear-response theory framework: $u = a_+ e^{i (\Vec{k} \cdot \Vec{x} - \omega t + \delta)} + a_- e^{- i (\Vec{k} \cdot \Vec{x} - \omega t + \delta)}$ with $a_\pm \in \mathbb{R}$.
Inserting this ansatz into the dynamical equation \eqref{eq-magnetic-GP}, we obtain the Bogoliubov equation
\begin{align}
\bigg(\frac{\omega^2}{c^2} \pm 2 \mu \omega\bigg) a_\pm = \Vec{k}^2 a_\pm + O(a_\pm^2),
\end{align}
giving the dispersion relation $\omega_\mp\up{H}$ and $\omega_\mp\up{NG}$ in Eq. \eqref{eq:dispersion-magnon} with arbitrary $a_\pm \neq 0$.
The gapful mode for $a_\mp\up{H} e^{\mp i (\Vec{k} \cdot \Vec{x} - \omega_\pm\up{H} t + \delta)}$ and NG mode for $a_\pm\up{NG} e^{ \pm i (\Vec{k} \cdot \Vec{x} - \omega_\pm\up{NG} t + \delta)}$ propagate in the direction parallel to $\Vec{k}$ for the upper sign and antiparallel to $\Vec{k}$ for the lower sign, respectively, 
where their chiralities are opposite to each other.
In the ultrarelativistic limit, $\omega_\pm\up{H} = \omega_\pm\up{NG} = \pm ck$ leads $a_\mp\up{H} \{ e^{\mp i (\Vec{k} \cdot \Vec{x} - \omega_\pm\up{H} t + \delta)} + e^{\pm i (\Vec{k} \cdot \Vec{x} - \omega_\pm\up{NG} t + \delta)} \} = 2 a_\mp\up{H} \cos(\Vec{k} \cdot \Vec{x} \mp ck t + \delta)$ with $a_\mp\up{H} = a_\pm\up{NG}$ and $a_\mp\up{H} \{ e^{\mp i (\Vec{k} \cdot \Vec{x} - \omega_\pm\up{H} t + \delta)} - e^{\pm i (\Vec{k} \cdot \Vec{x} - \omega_\pm\up{NG} t + \delta)} \} = \mp 2 i a_\mp\up{H} \sin(\Vec{k} \cdot \Vec{x} \mp ck t + \delta)$ with $a_\mp\up{H} = - a_\pm\up{NG}$, which give purely real and imaginary modes.
These results completely agree with those obtained in the low-energy effective theory.

\section{NG modes localized around solitons}\label{sec:soliton}

\subsection{Kelvon and dilaton-magnon of a Skyrmion}

We start from the $\mathbb{C}P^1$ Lagrangian density $\mathcal{L}$ in Eq. \eqref{eq-uniform-CP1}.
Instead of the uniform $u$, we consider 
a straight Skyrmion-line solution 
\cite{Polyakov:1975yp} 
extending along the $z$ axis,
\begin{align}
u\sub{s}(x,y,z) = \frac{\bar{r} e^{i (\bar{\theta} + \theta)}}{R\sub{s} + R},\quad
\bar{r} = \sqrt{(x - X)^2 + (y - Y)^2}, \quad
\bar{\theta} = \tan^{-1} \frac{y - Y}{x - X}, \label{eq:skyrmion-ansatz}
\end{align}
as the static solution.
Here, $R\sub{s} \in \mathbb{R}^+$ is the characteristic size of the Skyrmion line, and $X$, $Y \in \mathbb{R}$, $0 \leq \theta \leq 2 \pi$, and $R \in \mathbb{R}$ are the translational, phase, and dilatation moduli of the Skyrmion line, respectively.

\subsubsection{Low-energy effective theory}

To discuss the low-energy dynamics of the Skyrmion line, 
we use the moduli approximation 
\cite{Manton:1981mp}; we introduce the $z$ and $t$ dependences of these four moduli and integrate the Lagrangian density in the $xy$-plane with radius $L$:
\begin{align}
\begin{split}
&\int_{-L}^L dx\: \int_{- \sqrt{L^2 - x^2}}^{\sqrt{L^2 - x^2}} dy\: \mathcal{L}
= - 2 \pi + \pi \bigg\{ \frac{\dot{X}^2 + \dot{Y}^2}{c^2} - (X_z^2 + Y_z^2) + 2 \mu (\dot{X} Y - X \dot{Y}) \bigg\} \\
& + 2 \pi \log\bigg(\frac{L}{R\sub{s}}\bigg) \bigg\{ \frac{\dot{R}^2 + R\sub{s}^2 \dot{\theta}^2}{c^2} - (R_z^2 + R\sub{s}^2 \theta_z^2) + 2 \mu (R\sub{s}^2 \dot{\theta} + 2 R\sub{s} R \dot{\theta}) \bigg\}
+ O((X,Y,\theta,R)^3).
\end{split}
\end{align}
The first $2 \pi$ term in the right-hand side is the tension (the energy per unit length) of the Skyrmion line.
The low-energy dynamics of $X$, $Y$, $\theta$, and $R$ derived from the Euler--Lagrange equation becomes
\begin{subequations}
\begin{align}
& \ddot{X} = c^2 X_{zz} - 2 \mu c^2 \dot{Y}, \quad
\ddot{Y} = c^2 Y_{zz} + 2 \mu c^2 \dot{X}, \label{eq-skyrmion-kelvon} \\
& R\sub{s} \ddot{\theta} = c^2 R\sub{s} \theta_{zz} - 2 \mu c^2 \dot{R}, \quad
\ddot{R} = c^2 R_{zz} + 2 \mu c^2 R\sub{s} \dot{\theta}. \label{eq-skyrmion-dilaton}
\end{align}
\end{subequations}
Equation~\eqref{eq-skyrmion-kelvon} has the same form as Eq.~\eqref{eq-magnon-dynamics} with rewriting $\nabla \to \partial_z$, $\alpha_1 \to X$, and $\alpha_2 \to Y$.
Typical solutions $X = X_0 \cos(k z - \omega t + \delta)$ and $Y = Y_0 \sin(k z - \omega t + \delta)$ are therefore the same as those in Eq.~\eqref{eq:dispersion-magnon}.
As long as $\mu \neq 0$, the two translational moduli $X$ and $Y$ couple to each other forming gapless and gapful helical modes with $X_0 = Y_0$ and $X_0 = - Y_0$ propagating along the $z$ direction helically and antihelically.
The former is nothing but a helical Kelvin wave or a helical kelvon if quantized 
as a particle, while the latter may be called a 
``massive helical kelvon.''
The moduli fields $\theta$ and $R$ have the solution $\theta = \theta_0 \cos(k z - \omega t + \delta)$, $R = R_0 \sin(k z - \omega t + \delta)$ with the dispersion shown in Eq. \eqref{eq:dispersion-magnon}.
The 
phase and dilatation moduli $\theta$ and $R$ couple to each other forming gapless and gapful modes with $\theta_0 = R_0 / R\sub{s}$ and $\theta_0 = - R_0 / R\sub{s}$ propagating along 
the $z$ direction helically and antihelically. 
We called the former a ``dilaton-magnon'' \cite{Kobayashi:2014eqa}, and 
the latter may be called a ``massive dilaton-magnon.''  
In the ultrarelativistic limit $\mu \to 0$, the four modes for $X$, $Y$, $\theta$, and $R$ propagate independently of each other with the linear dispersion $c k$ as wavy kelvons for $X$ and $Y$, $U(1)$ magnon for $\theta$, and dilaton for $R$.

Here, we note that the dilatation symmetry is not the symmetry of the Lagrangian density \eqref{eq-uniform-CP1} but the symmetry of the stationary state of the dynamical equation for the Lagrangian, and the dilaton is not the NG mode but a so-called quasi-NG (QNG) mode \cite{Nitta:2014jta}, 
while wavy kelvons and phonons are NG modes.
The dilaton-magnon is also regarded as coupled NG--QNG mode, while a helical kelvon is a coupled NG mode.
In the nonrelativistic limit $c \to \infty$, 
the massive helical kelvon and massive dilaton-magnon 
disappear because of the divergent dispersion relation.

In Ref.~\cite{Kobayashi:2014xua},  
NG modes localized around a domain wall in ferromagnets 
with one easy axis were studied. 
The model is a nonrelativistic version of 
the massive ${\mathbb C}P^1$ model often studied in 
the supersymmetric context 
\cite{Abraham:1992vb}.
The domain wall breaks the translational symmetry transverse to 
the wall as well as the internal $U(1)$ symmetry.
There appear an associated ripple mode and $U(1)$ NG modes, 
coupled to each other.
The interpolation between ultrarelativistic and 
nonrelativistic theories is parallel to 
the case of a Skyrmion line.

\subsubsection{Linear-response theory}

We consider the dynamics of the Kelvin wave and dilaton-magnon in the linear-response theory framework: $u = u\sub{s}(R = \theta = X = Y = 0) + a_+ e^{i (\Vec{k} \cdot \Vec{x} - \omega t + \delta)} + a_- e^{- i (\Vec{k} \cdot \Vec{x} - \omega t + \delta)}$.
Inserting this ansatz into the dynamical equation \eqref{eq-magnetic-GP}, we obtain the Bogoliubov--de Gennes equation:
\begin{align}
\bigg( \frac{\omega^2}{c^2} \pm 2 \mu \omega \bigg) a_\pm = \bigg\{ (k^2 - \nabla_{\Vec{r}}^2) + \frac{4 (r \partial_r \pm i \partial_\theta)}{r^2 + R\sub{s}^2} \bigg\} a_\pm + O(a_\pm^2),
\end{align}
where, $\nabla_{\Vec{r}} = (\partial_x, \partial_y)$ denotes the derivative in the $xy$ plane.
By expanding $a_\pm$ as $a_\pm = \sum_l a_{\pm,l} e^{i l \theta}$, we obtain
\begin{align}
\bigg( \frac{\omega^2}{c^2} \pm 2 \mu \omega \bigg) a_{\pm,l} = \bigg\{ (k^2 - \partial_r^2 - \partial_r / r + l^2 / r^2) + \frac{4 (r \partial_r \mp l)}{r^2 + R\sub{s}^2} \bigg\} a_{\pm,l} + O(a_\pm^2).
\end{align}
There are two characteristic solutions: $a_{\pm,0} \propto 1$ with $l = 0$ and $a_{\pm,1} \propto r / R\sub{s}$ with $l = 1$ with the dispersion relation shown in Eq. \eqref{eq:dispersion-magnon}.
As long as $\mu \neq 0$, there are a gapless NG mode with $\omega_\pm\up{NG}$ and gapful Higgs mode with $\omega_\pm\up{H}$, and
the solution becomes
\begin{align}
\begin{split}
u_{\pm,0}\up{NG} &= \frac{r e^{i \theta}}{R\sub{s}} - X_0 e^{\pm i ( k z - \omega_\pm\up{NG} t + \delta)}, \quad
u_{\pm,0}\up{H} = \frac{r e^{i \theta}}{R\sub{s}} - X_0 e^{\mp i ( k z - \omega_\pm\up{H} t + \delta)},
\end{split}
\end{align}
for $l = 0$ and
\begin{align}
\begin{split}
u_{1}\up{NG} &= \frac{r e^{i \theta}}{R\sub{s}} + \frac{i \theta_0 r e^{\pm i ( k z - \omega_\pm\up{NG} t + \delta)}}{R\sub{s}}, \quad
u_{1}\up{H} = \frac{r e^{i \theta}}{R\sub{s}} + \frac{i \theta_0 r e^{\mp i ( k z - \omega_\pm\up{H} t + \delta)}}{R\sub{s}},
\end{split}
\end{align}
for $l = 1$.
$u_{\pm,0}\up{NG}$, $u_{\pm,0}\up{H}$, $u_{\pm,1}\up{NG}$, and $u_{\pm,1}\up{H}$ are equivalent to the solution \eqref{eq:skyrmion-ansatz} with moduli $X$, $Y$, $\theta$, and $R$ for helical kelvon, massive helical kelvon, dilaton-magnon, and massive dilaton-magnon, respectively, in the first order of $X_0$ and $\theta_0$.
In the ultrarelativistic limit $\mu \to 0$, both $\omega_\pm\up{NG}$ and $\omega_\pm\up{H}$ have linear dispersion relations $\pm ck$, giving wavy kelvons as the linear combination of $u_{\pm,0}\up{NG}$ and $u_{\pm,0}\up{H}$ and the $U(1)$ magnon and dilaton as the linear combination of $u_{\pm,1}\up{NG}$ and $u_{\pm,1}\up{H}$.

We shortly note other solutions having the same dispersions $\omega_\pm\up{NG}$ and $\omega_\pm\up{H}$.
With $l = 0$ and $l = 1$, solutions $a_{\pm,0}$ and $a_{\pm,1}$ have their anomalous pairs $a_{\pm,0} \propto (r / R\sub{s})^4 + 4 (r / R\sub{s})^2 + 4 \log(r / R\sub{s})$ and $a_{\pm,1} \propto (r / R\sub{s})^3 - (R\sub{s} / r) + 8 (r / R\sub{s}) \log(r / R\sub{s})$, which are the modes changing the Skyrmion charge of the total volume in which we are not interested.
For higher $l \geq 2$, there are also solutions $a_{\pm,l\geq 2} \propto (r / R\sub{s})^l$.
They do not change the state around the Skyrmion at the center but change the bulk state far from the Skyrmion, giving bulk magnons with $\omega_\pm\up{NG}$ and a bulk massive magnon with $\omega_\pm\up{H}$ propagating along arbitrary directions.
For lower $l < 0$, solutions $a_{\pm,l < 0} \propto (R\sub{s} / r)^{-l}$ give the Skyrmion-splitting modes from 1 Skyrmion at the center with the charge $+1$ to 1 Skyrmion with the charge $-l$ at the center and $(l+1)$ Skyrmions with the charge $+1$ around the center.
As well as the dilaton, these Skyrmion-splitting modes do not come from the symmetry of the Lagrangian density, and can be regarded as 
QNG modes and their massive partners.

\section{Summary and Discussion}\label{sec:summary}
In summary, we have revealed how 
relativistic and nonrelativistic NG modes 
are interpolated.  
We have found that type-I and type-II NG modes 
in the interpolating Lagrangians 
with the Lorentz invariance are accompanied with 
a gapful Higgs mode and gapful chiral NG partner, respectively.
In the ultrarelativistic limit, 
the type-I NG and Higgs partner remain, 
and the type-II NG mode and 
gapful NG partner become two type-I NG modes.
In the nonrelativistic limit, 
the accompanied gapful modes become infinitely massive,  disappearing from the spectrum. 
In the whole region, the commutation relation holds 
consistently, showing that the Lorentz invariance 
does not forbid type-II NG modes.

While we have studied a kelvon localized around a Skyrmion line 
in ferromagnets, 
we have not studied a kelvon localized around a vortex line  
in scalar BECs. 
In the latter case, 
the dispersion relation of the kelvon depends on 
the transverse system size.
When the size is finite, the dispersion relation is quadratic, 
but it is not an integer anymore for infinite system size.
Recently, the interpolating formula of the 
dispersion relation of the kelvon for an arbitrary transverse 
system size  was obtained in Ref.~\cite{Takahashi:2015caa}.
Extending to the case of a kelvon of a vortex 
in an arbitrary system size for interpolating relativistic and 
nonrelativistic systems is an interesting future work 
that would be important for the Mott insulator transition 
point in BECs in the optical lattice.

One of the related topics is the relaxation dynamics of topological defects generated through the Kibble--Zurek mechanism
after the temperature quench.
It has been predicted \cite{Bray:1994} that the relaxation dynamics is universal and dependent only on
a small number of factors such as conserved quantities, external currents, viscosity, and off-criticality.
For one of the future problems in this topic, we can consider the dependence of the relaxation dynamics on the dispersion relation
of NG modes excited along the topological defects in the interpolation between relativistic and nonrelativistic regions, which is expected to be an essential problem in the phase-ordering dynamics of $U(1)$ bosons with the finite chemical potential \cite{Laguna:1998}. 

Finally, let us mention 
quantum corrections of NG modes in lower dimensions.
In nonrelativistic limit, 
 type-II NG modes remain gapless under nonperturbative 
quantum corrections  \cite{Nitta:2013wca}, 
as opposed to type-I NG modes that become gapful 
to be consistent with the Coleman--Mermin--Wargner theorem.
It is interesting to see whether quantum corrections 
give gaps to type-II NG modes in the intermediate region.

\section*{Acknowledgments}
We thank the anonymous referees of 
our previous paper \cite{Kobayashi:2014xua} 
for the helpful suggestions and comments.
We also thank Aron J. Beekman and I. Danshita for useful discussions and comments.
The work of M.N. is supported in part
by the Japan Society for the Promotion of Science (JSPS)
Grant-in-Aid for Scientific Research
(KAKENHI Grant No. 25400268),
and
by a Grant-in-Aid for Scientific Research on Innovative Areas
``Topological Materials Science"
 (KAKENHI Grant No. 15H05855),
``Nuclear Matter in Neutron Stars Investigated by Experiments and
Astronomical Observations"
(KAKENHI Grant No. 15H00841),
and
``Topological Quantum Phenomena"
(KAKENHI Grant No. 25103720)
from the Ministry of Education, Culture, Sports, Science,
and Technology (MEXT) of Japan.
The work of M.N. is also supported in part by
the MEXT-Supported Program for the Strategic Research Foundation
at Private Universities ``Topological Science" (Grant No. S1511006).
The work of M.K. is
supported in part by Grant-in-Aid for Scientific Research
(Grant No. 26870295); by a Grant-in-Aid for Scientific Research on
Innovative Areas “Fluctuation \& Structu'' (Grant No. 26103519)
from the Ministry of Education, Culture, Sports, Science,
and Technology of Japan; and by the JSPS Core-to-Core program
``Non-Equilibrium Dynamics of Soft-Matter and Information."


\begin{appendix}

\section{Spinor BEC}\label{sec:spinor}

Here, we discuss a ferromagnetic $F=1$ 
spinor BEC which contains both type-I and -II 
NG modes simultaneously. 
We see that it contains further a pair of gapful modes.  
The interpolating Lagrangian is 
\begin{align}
\mathcal{L} = \frac{1}{c^2} |\partial_t \psi|^2 + i \mu (\psi^\dagger \partial_t \psi - \partial_t \psi^\dagger \psi) - |\nabla \psi|^2 - \frac{g_0}{2} \bigg\{|\psi|^2 - \frac{(g_0 - g_1) \rho}{g_0} \bigg\}^2 + \frac{g_1}{2} (\psi^\dagger \hat{\Vec{F}} \psi)^2, \label{eq-spinor-Lagrangian}
\end{align}
where $\psi = (\psi_{1}, \psi_0, \psi_{-1})^T$ is the  
three-component (spinor-1) complex scalar fields and $\hat{\Vec{F}}$ is 
the triplet of the 3 by 3 $SO(3)$ generators (spin-1 spin matrices).
Besides the Lorentz transformation given in Eq.~\eqref{eq:Lorentz} for all components $\psi_{\pm1}$ and $\psi_0$,  this Lagrangian is invariant under the shift of the overall phase $\psi \to \psi e^{i \theta}$ and the $SO(3)$ spin rotation $\psi \to \psi e^{- i \hat{\Vec{F}} \cdot \Vec{s}}$.
When the two coupling constants $g_0$ and $g_1$ satisfy $g_0 > g_1 \geq 0$, there exists a stable and static solution $\psi\sub{g} = (\sqrt{\rho},0,0)^T$ for $\Vec{F} = \psi^\dagger \hat{\Vec{F}} \psi = (0,0,\rho) $ as the ground state.


As well as the previous examples, we consider the following low-energy excited state:
\begin{align}
\psi = \sqrt{\rho} ((1 + f_1) e^{i \theta_1}, \alpha_0 + i \beta_0, \alpha_{-1} + i \beta_{-1})^T, \label{eq-spinor-low-energy}
\end{align}
where $f_1$ and $\theta_1$ are fluctuations of the amplitude and phase of the first component of $\psi$ and $\alpha_m$ 
and $\beta_m$ are the real and imaginary parts of fluctuations of 
the $m$th component ($m = 0, -1$) of $\psi$. 
Inserting Eq.~\eqref{eq-spinor-low-energy} into Eq.~\eqref{eq-spinor-Lagrangian}, we get the effective Lagrangian
\begin{align}
\begin{split}
\frac{\mathcal{L}}{\rho} &= \frac{ \dot{f}_1^2 + \dot{\theta}_1^2 + \dot{\alpha}_0^2 + \dot{\beta}_0^2 + \dot{\alpha}_{-1}^2 + \dot{\beta}_{-1}^2}{c^2} \\
&\quad - 2 \mu \{ (1 + 2 f_1) \dot{\theta}_1 + \alpha_0 \dot{\beta}_0 - \dot{\alpha}_0 \beta_0 + \alpha_{-1} \dot{\beta}_{-1} - \dot{\alpha}_{-1} \beta_{-1} \} \\
&\quad - ( |\nabla f_1|^2 + |\nabla \theta_1|^2 + |\nabla \alpha_0|^2 + |\nabla \beta_0|^2 + |\nabla \alpha_{-1}|^2 + |\nabla \beta_{-1}|^2) \\
&\quad - 2 (g_0 - g_1) \rho f_1^2 - 2 g_1 \rho (\alpha_{-1}^2 + \beta_{-1}^2) + O((f_1, \theta_1, \alpha_0, \beta_0, \alpha_{-1}, \beta_{-1})^3).
\end{split}
\end{align}
The low-energy dynamics becomes
\begin{subequations}
\begin{align}
& \frac{\ddot{f}_1}{c^2} + 2 \mu \dot{\theta}_1 - \nabla^2 f_1 + 2 (g_0 - g_1) \rho f_1 = 0, \quad
\frac{\ddot{\theta}_1}{c^2} - 2 \mu \dot{\theta}_1 - \nabla^2 \theta_1 = 0, \label{eq-spinor-phonon} \\ 
& \frac{\ddot{\alpha}_{0}}{c^2} + 2 \mu \dot{\beta}_0 - \nabla^2 \alpha_0 = 0, \quad
\frac{\ddot{\beta}_{0}}{c^2} - 2 \mu \dot{\alpha}_0 - \nabla^2 \beta_0 =0, \label{eq-spinor-magnon} \\ 
& \frac{\ddot{\alpha}_{-1}}{c^2} + 2 \mu \dot{\beta}_{-1} - \nabla^2 \alpha_{-1} + 2 g_1 \rho \alpha_1 = 0, \quad
\frac{\ddot{\beta}_{-1}}{c^2} - 2 \mu \dot{\alpha}_{-1} - \nabla^2 \beta_{-1} + 2 g_1 \rho \beta_1 =0. \label{eq-spinor-gap} 
\end{align}
\end{subequations}
Equation~\eqref{eq-spinor-phonon} has the same form as Eq.~\eqref{eq-scalar-Euler-Lagrange} with rewriting $f \to f_1$, $\theta \to \theta_1$, and $g \to g_0 - g_1$.
Therefore, as long as $\mu \neq 0$, dynamics of $f_1$ and $\theta_1$ are coupled as $f_1 = f_{10} \cos(\Vec{k} \cdot \Vec{x} - \omega t + \delta)$ and $\theta_1 = \theta_{10} \sin(\Vec{k} \cdot \Vec{x} - \omega t + \delta)$ with the dispersions shown in 
Eqs.~\eqref{eq:higgs-phonon} and \eqref{eq-scalar-amplitude}: one gapful Higgs mode with $\omega_\pm\up{H}$ and one type-I gapless NG mode with $\omega_\pm\up{NG}$.
In the relativistic limit, two dynamics of $f_1$ and $\theta_1$ are independent of each other giving a Higgs mode for $f_1$ and 
a type-I NG mode for $\theta_1$.
In the nonrelativistic limit, the Higgs mode vanishes with a diverging spectrum.

Equation~\eqref{eq-spinor-magnon} has the same form as Eq.~\eqref{eq-magnon-dynamics}, and the solutions $\alpha_0 = \alpha_{00} \cos(\Vec{k} \cdot \Vec{x} - \omega t + \delta)$ and $\beta_0 = \beta_{00} \sin(\Vec{k} \cdot \Vec{x} - \omega t + \delta)$ exactly behave as $\alpha$ and $\beta$: one Higgs mode with $\omega = \omega_\pm\up{H}$ and $\beta_{00} = \mp \alpha_{00}$ 
and one type-II NG mode with $\omega = \omega_\pm\up{NG}$ and $\beta_{00} = \pm \alpha_{00}$ having opposite chiralities as long as $\mu \neq 0$.
In the ultrarelativistic limit, the Higgs and type-II NG modes are degenerated, giving rise to two type-I NG modes.
In the nonrelativistic limit, 
the Higgs mode vanishes with a diverging spectrum.
The modes $\alpha_0$ and $\beta_0$ can be considered as fluctuations of the spin rotation around $\Vec{F} = (0,0,\rho)$.
Fluctuations for $F_x$ and $F_y$ can be written as $e^{- i \hat{F}_y s} \psi\sub{g} = \sqrt{\rho} (1, s / \sqrt{2}, 0)^T + O(s^2) $ and $e^{i \hat{F}_x s} \psi\sub{g} = \sqrt{\rho} (1, i s / \sqrt{2}, 0)^T + O(s^2)$.
The real and imaginary parts $\alpha_0$ and $\beta_0$ therefore correspond to fluctuations of $F_x$ and $F_y$, 
respectively, which is consistent with $\alpha$ and $\beta$ in Eq. \eqref{eq-magnon-effective-Lagrangian} as fluctuations of $n_1$ and $n_2$.

For Eq.~\eqref{eq-spinor-gap}, typical solutions are $\alpha_{-1} = \alpha_{-10} \cos(\Vec{k} \cdot \Vec{x} - \omega t + \delta)$ and $\beta_{-1} = \beta_{-10} \sin(\Vec{k} \cdot \Vec{x} - \omega t + \delta)$ as long as $\mu \neq 0$ with the dispersion
\begin{align}
\begin{split}
\omega\up{G1}_\pm &= \pm c \Big( \sqrt{k^2 + \mu^2 c^2 + 2 g_1 \rho} + \mu c \Big) \\
&= \pm c \bigg( \sqrt{ \mu^2 c^2 + 2 g_1 \rho } + \mu c + \frac{k^2}{2 \sqrt{ \mu c^2 + 2 g_1 \rho}} \bigg) + O(k^4), \quad \alpha\up{H}_{-10\pm} = \mp \beta\up{H}_{-10\pm}, \\
\omega\up{G2}_\pm &= \pm c \Big( \sqrt{k^2 + \mu^2 c^2 + 2 g_1 \rho} - \mu c \Big) \\
&= \pm c \bigg( \sqrt{ \mu^2 c^2 + 2 g_1 \rho } - \mu c + \frac{k^2}{2 \sqrt{ \mu c^2 + 2 g_1 \rho}} \bigg) + O(k^4), \quad \alpha\up{H}_{-10\pm} = \pm \beta\up{H}_{-10\pm}.
\label{eq:dispersion-polar}
\end{split}
\end{align}
Being different from $\omega_\pm\up{NG}$ and $\omega_\pm\up{H}$ for $\alpha_0$ and $\beta_0$, both $\omega_\pm\up{G1}$ and $\omega_\pm\up{G2}$ are gapful as long as $g_1 > 0$.
In the nonrelativistic limit $c \to \infty$, $\omega_\pm\up{G1}$ diverges as well as $\omega_\pm\up{H}$, and only the gapful mode with $\omega_\pm\up{G2}$ survives.
In the ultrarelativistic limit $\mu \to 0$, $\alpha_{-1}$ and $\beta_{-1}$ are independent of each other with the dispersion
\begin{align}
\omega_\pm\up{G1} = \omega_\pm\up{G2} = \pm c \bigg( \sqrt{2 g_1 \rho} + \frac{k^2}{2 \sqrt{2 g_1 \rho}} \bigg) + O(k^4),
\end{align}
which remains gapful.
Defining new operators 
\begin{align}
\hat{F}_{p1} = \begin{pmatrix} 0 & 0 & 1 \\ 0 & 0 & 0 \\ 1 & 0 & 0 \end{pmatrix}, \quad
\hat{F}_{p2} = \begin{pmatrix} 0 & 0 & -i \\ 0 & 0 & 0 \\ i & 0 & 0 \end{pmatrix},
\end{align}
we can write the modes corresponding to $\alpha_{-1}$ and $\beta_{-1}$ as $e^{- i \hat{F}_{p2} s} \psi\sub{g} = \sqrt{\rho} (1, 0, s)^T + O(s^2)$ and $e^{i \hat{F}_{p1} s} \psi\sub{g} = \sqrt{\rho} (1, 0, i s)^T + O(s^2)$.
Because we can obtain the nonmagnetic polar state $\psi\sub{p} = \sqrt{\rho} (1,0,e^{i \delta})^T / \sqrt{2}$ with $\hat{F}_{p1}$ and $\hat{F}_{p2}$ as $e^{- i ( \hat{F}_{p2} \cos\delta - \hat{F}_{p1} \sin\delta) \pi /4} \psi\sub{g} = \psi\sub{p}$, we can regard $\alpha_{-1}$ and $\beta_{-1}$ as the fluctuation from the ferromagnetic state to the polar state.

In the case of $g_1 = 0$, the number of NG modes changes as follows.
In this case, two dispersions $\omega_\pm\up{G1}$ and $\omega_\pm\up{G2}$ become equivalent to $\omega_\pm\up{H}$ and $\omega_\pm\up{NG}$ for $\alpha_0$ and $\beta_0$ respectively, and $\alpha_{-1}$ and $\beta_{-1}$ also contribute to the type-II NG and Higgs modes.
This is a consequence  of the fact that the symmetry of the Lagrangian is enlarged from $U(1) \times SO(3)$ to $U(3)$ and broken generators for $\psi\sub{g}$ included in $u(3)$ are $\hat{F}_{x,y,z,p1,p2}$, which have been just considered above.

We finally refer to the linear-response theory, which gives the same results as those from the low-energy effective theory as well as other examples in the main part.
The Bogoliubov equation can be obtained by 
substituting the fluctuations $\psi \to \psi\sub{g} + u e^{i (\Vec{k} \cdot \Vec{x} - \omega t + \delta)} + v^\ast e^{- i (\Vec{k} \cdot \Vec{x} - \omega t + \delta)}$ to the original Lagrangian \eqref{eq-spinor-Lagrangian},
\begin{align}
\begin{pmatrix}
\omega^2 / c^2 + 2 \mu \omega - k^2 - F & - G \\
- G & \omega^2 / c^2 - 2 \mu \omega - k^2 - F
\end{pmatrix}
\begin{pmatrix} u \\ v \end{pmatrix} + O((u,v)^2)= 0
\end{align}
where $F$ and $G$ are given in 
the ferromagnetic ground state $\psi\sub{g}$ as 
\begin{align}
F = \rho \begin{pmatrix}
g_0 - g_1 & 0 & 0 \\
0 & 0 & 0 \\
0 & 0 & 2 g_1
\end{pmatrix} ,\ \quad
G = \rho \begin{pmatrix}
g_0 - g_1 & 0 & 0 \\
0 & 0 & 0 \\
0 & 0 & 0
\end{pmatrix}.
\end{align}
We obtain the same dispersion relations as those discussed in the low-energy effective theory.

\end{appendix}


\begin{thebibliography}{100}


\bibitem{Nielsen:1975hm} 
  H.~B.~Nielsen and S.~Chadha,
  ``On How to Count Goldstone Bosons,''
  Nucl.\ Phys.\ {\bf B105}, 445 (1976).

\bibitem{Nambu:2004yia} 
  Y.~Nambu,
  ``Spontaneous breaking of Lie and current algebras,''
  J.\ Stat.\ Phys.\  {\bf 115}, 7 (2004).

\bibitem{Watanabe:2011ec} 
  H.~Watanabe and T.~Brauner,
  ``Number of Nambu--Goldstone bosons and its relation to charge densities,''
  Phys.\ Rev.\ D {\bf 84}, 125013 (2011)
  [arXiv:1109.6327 [hep-ph]].

\bibitem{Watanabe:2012hr} 
  H.~Watanabe and H.~Murayama,
  ``Unified Description of Nambu--Goldstone Bosons without Lorentz Invariance,''
  Phys.\ Rev.\ Lett.\  {\bf 108}, 251602 (2012)
  [arXiv:1203.0609 [hep-th]].


\bibitem{Watanabe:2014fva} 
  H.~Watanabe and H.~Murayama,
  ``Effective Lagrangian for Nonrelativistic Systems,''
  Phys.\ Rev.\ X {\bf 4}, 031057 (2014)
  [arXiv:1402.7066 [hep-th]].

\bibitem{Hidaka:2012ym} 
  Y.~Hidaka,
  ``Counting Rule for Nambu--Goldstone Modes in Nonrelativistic Systems,''
  Phys.\ Rev.\ Lett.\  {\bf 110}, 091601 (2013)
  [arXiv:1203.1494 [hep-th]].

\bibitem{Takahashi:2014vua} 
  D.~A.~Takahashi and M.~Nitta,
  ``Counting rule of Nambu--Goldstone modes for internal and spacetime symmetries: Bogoliubov theory approach,''
  Ann.\ Phys.\  {\bf 354}, 101 (2015)
  [arXiv:1404.7696 [cond-mat.quant-gas]].

\bibitem{Kobayashi:2014xua} 
  M.~Kobayashi and M.~Nitta,
  ``Nonrelativistic Nambu--Goldstone Modes Associated with Spontaneously Broken Space-Time and Internal Symmetries,''
  Phys.\ Rev.\ Lett.\  {\bf 113}, 120403 (2014)
  [arXiv:1402.6826 [hep-th]].


\bibitem{Watanabe:2014pea} 
  H.~Watanabe and H.~Murayama,
  ``Noncommuting Momenta of Topological Solitons,''
  Phys.\ Rev.\ Lett.\  {\bf 112}, 191804 (2014)
  [arXiv:1401.8139 [hep-th]].

\bibitem{Kobayashi:2014eqa} 
  M.~Kobayashi and M.~Nitta,
  ``Non-relativistic Nambu--Goldstone modes propagating along a skyrmion line,''
  Phys.\ Rev.\ D {\bf 90}, 025010 (2014)
  [arXiv:1403.4031 [hep-th]].

\bibitem{Kobayashi:2013gba} 
  M.~Kobayashi and M.~Nitta,
  ``Kelvin modes as Nambu--Goldstone modes along superfluid vortices and relativistic strings: Finite volume size effects,''
  Prog. Theor. Exp. Phys. {\bf 2014}, 021B01 (2014)
  [arXiv:1307.6632 [hep-th]].


\bibitem{Takeuchi:2013mwa} 
  H.~Takeuchi and K.~Kasamatsu,
  ``Nambu--Goldstone modes in segregated Bose--Einstein condensates,''
  Phys.\ Rev.\ A {\bf 88}, 043612 (2013)
  [arXiv:1309.3224 [cond-mat.quant-gas]].

\bibitem{Watanabe:2014zza} 
  H.~Watanabe and H.~Murayama,
  ``Nambu--Goldstone bosons with fractional-power dispersion relations,''
  Phys.\ Rev.\ D {\bf 89}, 101701 (2014)
 [arXiv:1403.3365 [hep-th]].

\bibitem{Takahashi:2015caa} 
  D.~A.~Takahashi, M.~Kobayashi and M.~Nitta,
  ``Nambu--Goldstone modes propagating along topological defects: Kelvin and ripple modes from small to large systems,''
  Phys.\ Rev.\ B {\bf 91}, 184501 (2015)
  [arXiv:1501.01874 [cond-mat.other]].

\bibitem{Higashikawa:2015}
  S.~Higashikawa and M.~Ueda,
  ``$\mu$-Symmetry breaking : Algebraic approach to finding building blocks of quantum many-body systems,''
  [arXiv:1504.0811 [cond-mat.quant-gas]].

\bibitem{Hidaka:2014fra} 
  Y.~Hidaka, T.~Noumi and G.~Shiu,
  ``Effective field theory for spacetime symmetry breaking,''
  [arXiv:1412.5601 [hep-th]].

\bibitem{Hama:2011rt} 
  Y.~Hama, T.~Hatsuda and S.~Uchino,
  ``Higgs mechanism with type-II Nambu--Goldstone bosons at finite chemical potential,''
  Phys.\ Rev.\ D {\bf 83}, 125009 (2011)
  [arXiv:1102.4145 [hep-ph]].

\bibitem{Gongyo:2014sra} 
  S.~Gongyo and S.~Karasawa,
  ``Nambu--Goldstone bosons and the Higgs mechanism without Lorentz invariance: Analysis based on constrained-system theory,''
  Phys.\ Rev.\ D {\bf 90}, 085014 (2014)
  [arXiv:1404.1892 [hep-th]].

\bibitem{Watanabe:2014qla} 
  H.~Watanabe and H.~Murayama,
  ``Spontaneously broken non-Abelian gauge symmetries in nonrelativistic systems,''
  Phys.\ Rev.\ D {\bf 90}, 121703 (2014)
  [arXiv:1405.0997 [hep-th]].

\bibitem{Hayata:2014yga} 
  T.~Hayata and Y.~Hidaka,
  ``Dispersion relations of Nambu--Goldstone modes at finite temperature and density,''
  Phys.\ Rev.\ D {\bf 91}, 056006 (2015)
  [arXiv:1406.6271 [hep-th]].


\bibitem{Brauner:2014ata} 
  T.~Brauner and S.~Moroz,
  ``Topological interactions of Nambu--Goldstone bosons in quantum many-body systems,''
  Phys.\ Rev.\ D {\bf 90}, 121701 (2014)
  [arXiv:1405.2670 [hep-th]].


\bibitem{Nitta:2014jta} 
  M.~Nitta and D.~A.~Takahashi,
  ``Quasi-Nambu--Goldstone modes in nonrelativistic systems,''
  Phys.\ Rev.\ D {\bf 91}, no. 2, 025018 (2015)
  [arXiv:1410.2391 [hep-th]].

\bibitem{Parisi:1983}
  G.~Parisi,
  ``On complex probabilities,''
  Phys.\ Lett.\ {\bf 131B}, 393 (1983).

\bibitem{Aarts:2008rr} 
  G.~Aarts and I.~O.~Stamatescu,
  ``Stochastic quantization at finite chemical potential,''
  J. High Energy Phys. 09 (2008) 018
  [arXiv:0807.1597 [hep-lat]].


\bibitem{Altman:2002}
  E.~Altman and A.~Auerbach,
  ``Oscillating Superfludity of Boson in Optical Lattices,''
  Phys.\ Rev.\ Lett. {\bf 89}, 250404 (2002).

\bibitem{Altman:2008}
  S.~D.~Huber, B.~ Theiler, E.~Altman, and G.~Blatter,
  ``Amplitude Mode in the Quantum Phase Model,''
  Phys.\ Rev.\ Lett. {\bf 100}, 050404 (2008).

\bibitem{Pollet:2012}
  L.~Pollet and N.~Prokof'ev,
  ``Higgs Mode in a Two-Dimensiona Superfluid,''
  Phys.\ Rev.\ Lett. {\bf 109}, 010401 (2012).

\bibitem{Grass:2011}
  T.~D.~Grass, F.~E.~A.~dos Santos, and A.~Pelster,
  ``Real-time Ginzburg--Landau theory for bosons in optical lattices,''
  Laser\ Phys.\ {\bf 21}, 1459 (2011).

\bibitem{Krutitsky:2011}
  K.~V.~Krutitsky and P.~Navez,
  ``Excitation dynamics in a lattice Bose gas within the time-dependent Gutzwiller mean-field approach,''
  Phys.\ Rev.\ A\ {\bf 84}, 033602 (2011).

\bibitem{Nakayama:2015}
  T.~Nakayama, I.~Danshita, T.~Nikuni, and S.~Tsuchiya,
  ``Fano resonance through Higgs bound states in tunneling of Nambu--Goldstone modes,''
  [arXiv:1503.01516 [cond-mat.quant-gas]].

\bibitem{Bissbort:2011}
  U.~Bissbort, S.~G\"otze, Y.~Li, J.~Heinze, J.~S.~Krauser, M.~Weinberg, C.~Becker, K.~Sengstock, and W.~Hofstetter
  ``Detecting the Amplitude Mode of Stringly Interacting Lattice Bosons by Bragg Scattering,''
  Phys.\ Rev.\ Lett. {\bf 106}, 205303 (2011).

\bibitem{Liu:2015}
  B.~Liu, H.~Zhai, and S.~Zhang,
  ``Evolution of Higgs mode in a Fermion Superfluid with Tunable Interactions,''
  [arXiv:1502.00431].

\bibitem{Beekman:2014}
  A.~J.~Beekman,
  ``Criteria for the absence of quantum fluctuations after spontaneous symmetry breaking,''
  Ann. Phys. {\bf 361}, 461 (2015)
  [arXiv:1408.1691 [cont-mat.other]].


\bibitem{Polyakov:1975yp} 
  A.~A.~Belavin and A.~M.~Polyakov ,
  ``Metastable States of Two-Dimensional Isotropic Ferromagnets,''
  Pis'ma Zh. Eksp. Teor. Fiz. {\bf 22}, 503 (1975)
  [JETP Lett.\  {\bf 22}, 245 (1975)].

\bibitem{Manton:1981mp} 
  N.~S.~Manton,
  ``A Remark on the Scattering of BPS Monopoles,''
  Phys.\ Lett.\ {\bf 110B}, 54 (1982);
  M.~Eto, Y.~Isozumi, M.~Nitta, K.~Ohashi and N.~Sakai,
  ``Manifestly supersymmetric effective Lagrangians on BPS solitons,''
  Phys.\ Rev.\ D {\bf 73}, 125008 (2006)
  [hep-th/0602289].

\bibitem{Abraham:1992vb} 
  E.~R.~C.~Abraham and P.~K.~Townsend,
  ``Q kinks,''
  Phys.\ Lett.\ {\bf 291B}, 85 (1992);
  ``More on Q kinks: a (1+1)-dimensional analogue of dyons,''
  Phys.\ Lett.\ {\bf 295B}, 225 (1992);
  M.~Arai, M.~Naganuma, M.~Nitta and N.~Sakai,
  ``Manifest supersymmetry for BPS walls in N=2 nonlinear sigma models,''
  Nucl.\ Phys.\ {\bf B652}, 35 (2003)
  [hep-th/0211103]; 
  M.~Arai, M.~Naganuma, M.~Nitta and N.~Sakai,
  ``BPS wall in N=2 SUSY nonlinear sigma model with Eguchi-Hanson manifold,''
  In ``A garden of quanta* 299-325,'' edited by A. Arai {\it et. al}..
  (World Scientific, Singapore, 2003)
  [hep-th/0302028].


\bibitem{Bray:1994}
  A.~J.~Bray,
  ``Theory of phase-ordering kinetics''
  Adv. Phys. {\bf 43}, 357 (1994).


\bibitem{Laguna:1998}
A similar problem has been studied as the dependence of the relaxation dynamics with topological defects on
the viscosity in
P.~Laguna and W.~H.~Zurek,
  ``Critical dynamics of symmetry breaking: Quenches, dissipation, and cosmology''
  Phys. Rev. D {\bf 58}, 085021 (1998).
If we replace $\eta \dot{\varphi}$ with $i \eta \dot{\varphi}$ in the left-hand-side of Eq. (1) of the above paper, 
the problem just becomes the relaxation dynamics in the interpolating between relativistic and
nonrelativistic regions.


\bibitem{Nitta:2013wca} 
  M.~Nitta, S.~Uchino and W.~Vinci,
  ``Quantum Exact Non-Abelian Vortices in Non-relativistic Theories,''
  J. High Energy Phys. 09 (2014) 098
  [arXiv:1311.5408 [hep-th]].


\end{thebibliography}


\end{document}